\documentclass[a4paper]{article}

\usepackage{INTERSPEECH2016}
\usepackage{graphicx,amssymb,amsmath,bm,textcomp,subfigure,multirow,epstopdf,graphicx,tabularx}

\ninept

\title{DNN based Speaker Recognition on Short Utterances}

\makeatletter
\def\name#1{\gdef\@name{#1\\}}
\makeatother 
\name{{\em Ahilan Kanagasundaram$^{*+}$, David Dean$^{*}$, Sridha Sridharan$^{*}$ and Clinton Fookes$^{*}$}}
\address{Speech and Audio Research Laboratory$^{*}$  \\
	Queensland University of Technology, Brisbane, Australia$^{*}$ \\
	{\small \tt \{a.kanagasundaram, d.dean, s.sridharan, c.fookes\}@qut.edu.au}$^{*}$ \\
	Electrical \& Electronic Engineering, Faculty of Engineering$^{+}$ \\
	University of Jaffna, Jaffna, Sri Lanka$^{+}$\\
	{\small \tt ahilan@eng.jfn.ac.lk}$^{+}$
}

\begin{document}

\maketitle
\begin{abstract}
This paper investigates the effects of limited speech data in the context of speaker verification using deep neural network~(DNN) approach. Being able to reduce the length of required speech data is important to the development of speaker verification system in real world applications. The experimental studies have found that DNN-senone-based Gaussian probabilistic linear discriminant analysis~(GPLDA) system respectively achieves above 50\% and 18\% improvements in EER values over GMM-UBM GPLDA system on NIST 2010 coreext-coreext and truncated 15sec-15sec evaluation conditions. Further when GPLDA model is trained on short-length utterances~(30sec) rather than full-length utterances~(2min), DNN-senone GPLDA system achieves above 7\% improvement in EER values on truncated 15sec-15sec condition. This is because short length development i-vectors have speaker, session and phonetic variation and GPLDA is able to robustly model those variations. For several real world applications, longer utterances~(2min) can be used for enrollment and shorter utterances~(15sec) are required for verification, and in those conditions, DNN-senone GPLDA system achieves above 26\% improvement in EER values over GMM-UBM GPLDA systems.
\end{abstract}

\section{Introduction}\label{sec:Introduction}
In a typical speaker verification system, the significant amount of speech required for development, enrollment and verification in the presence of large inter-session variability has limited the widespread use of speaker verification technology in everyday applications. A number of recent studies in speaker verification design, including joint factor analysis~(JFA)~\cite{Vogt2008a,Vogt2008c,McLaren2010b}, i-vectors~\cite{Kanagasundaram2011} and probabilistic linear discriminant analysis~(PLDA)~\cite{Kanagasundaram2012a} has focused on reducing the amount of speech required during development, enrolment and verification while obtaining satisfactory performance. However, these studies have shown that the performance degrades considerably in short utterances for all common approaches. This paper will focus on whether a recently proposed deep neural network~(DNN) approach to speaker verification could form a suitable foundation for continuing research into short utterance speaker verification.

Recently, DNNs have been incorporated into i-vector-based speaker recognition systems using two main approaches: (1) A speech-based DNN is used to extract bottleneck~(BN) features from an inner layer restricted in dimensionality, and/or (2) senone-replacement, where the calculation of Baum-Welch statistics from a traditional Gaussian-mixture model is replaced by a speech-based DNN performing a similar function~\cite{McLarenLeiFerrer,SnyderGarcia-RomeroPovey2015,Garcia-RomeroZhangMcCreeEtAl2014,RichardsonReynoldsDehak2015,KennyGuptaStafylakisEtAl2014,RichardsonReynoldsDehak2015a}. BN speech features were first proposed for large vocabulary speech recognition in conjunction with conventional acoustic features, such as mel-frequency cepstral coefficients (MFCC)~\cite{HintonDengYuEtAl2012}, and have more recently shown similar improvement in language and speaker identification applications~\cite{McLarenLeiFerrer,RichardsonReynoldsDehak2015,RichardsonReynoldsDehak2015a}. Similarly, substitution of GMMs by DNNs have also worked well for speech recognition~\cite{HintonDengYuEtAl2012,DahlYuDengEtAl2012}, and have recently shown similar promise in language and speaker recognition~\cite{McLarenLeiFerrer,RichardsonReynoldsDehak2015a,SnyderGarcia-RomeroPovey2015}.

The main aim of this paper is to investigate the effect of only having short utterances~(15 sec) available for enrolment and verification for DNN-senone GPLDA speaker verification, which is already known to outperform GMM-senone systems with long utterances~(2 mins) for enrolment and verification~\cite{SnyderGarcia-RomeroPovey2015}. In addition to investigating short utterances during enrolment and verification, this paper will also investigate if GPLDA can benefit from having short utterances available during the development of the background GPLDA models, as has found to be the case in GMM-UBM GPLDA\cite{Kanagasundaram2012a}.

This paper is structured as follows: Section~\ref{sec:GMM-UBM based GPLDA system} outlines a GMM-UBM-based GPLDA system, and Section~\ref{sec:DNN senone posterior based GPLDA system} gives an overview of DNN-senone-based GPLDA system. Section~\ref{sec:DNN training} details DNN training approach. Section~\ref{sec:Methodology} details the methodology and experimental setup. The results and discussions are given in Section~\ref{sec:Results and Discussions} and Section~\ref{sec:Conclusion} concludes the paper.
\section{GMM-UBM-based GPLDA system}\label{sec:GMM-UBM based GPLDA system}
\begin{figure}
\centering
\subfigure[UBM-based i-vector extractor]{
\centerline{\includegraphics[width=7cm]{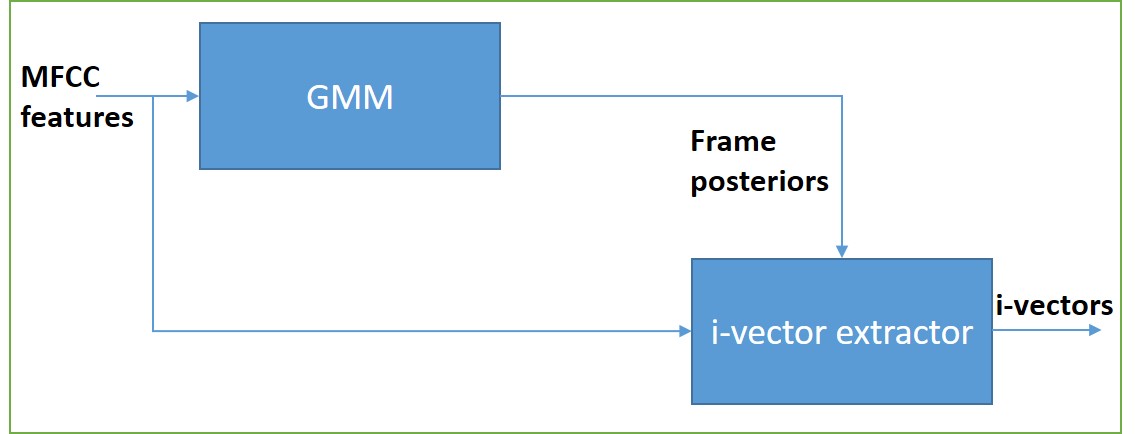}}
\label{fig:subfig1}
}
\subfigure[DNN senone posterior based i-vector extractor]{
\centerline{\includegraphics[width=7cm]{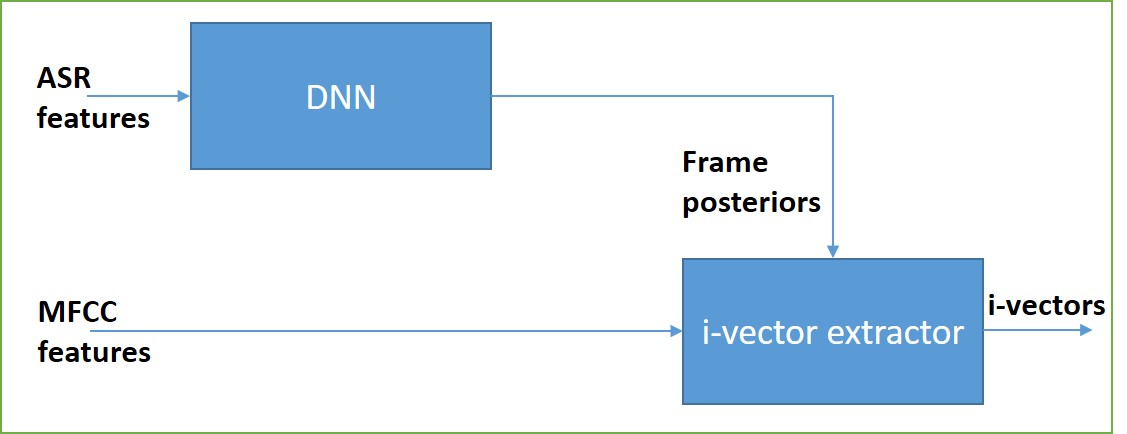}}
\label{fig:subfig2}
}
\label{fig:subfigureExample}
\caption{\emph{The block diagram of }, \subref{fig:subfig1} UBM-based, \emph{and} \subref{fig:subfig2} DNN senone posterior based i-vector extractor.}
\end{figure}

In the typical i-vector approach, the UBM is trained using expectation maximization~(EM) algorithm. The UBM is composed of $\textbf{C}$ Gaussian components. The UBM is then used to extract zeroth order, $\textbf{N}$,  first order, $\textbf{F}$, and second order, $\textbf{S}$ Baum-Welch statistics~(alternatively referred to as sufficient statistics). The zeroth order statistics are the total occupancies across an utterance for each GMM component and the first order statistics are the occupancy-weighted accumulations of feature vectors for each component. The extraction of i-vectors is based on the Baum-Welch zero-order and centralized first-order statistics. The statistics are calculated for a given utterance with respect to $C$ UBM components and $F$ dimension MFCC features. The i-vector for a given utterance can be extracted as follows~\cite{Dehak2010},
\begin{eqnarray}
\textbf{w} & = & (\textbf{I} + \textbf{T}^{T}\boldsymbol{\Sigma}^{-1}\textbf{N}\textbf{T})^{-1}\textbf{T}^{T}\boldsymbol{\Sigma}^{-1}\textbf{F},
\end{eqnarray}
where $\textbf{I}$ is a $CF \times CF$ identity matrix, $\textbf{N}$ is a diagonal matrix with $F \times F$ blocks $N_{c}\textbf{I}$~($c = 1,2,....C$), and the super-vector, $\textbf{F}$, is formed through the concatenation of the centralized first-order statistics. The covariance matrix, $\boldsymbol{\Sigma}$, represents the residual variability not captured by $\textbf{T}$. An efficient procedure of estimating the total-variability subspace, $\textbf{T}$, is described in~\cite{Kenny2008,Dehak2010}. The total-variability subspace, $\textbf{T}$, is learned in an unsupervised way on a large dataset to capture most of the speaker information. Figure~\ref{fig:subfigureExample}~(a) illustrates how i-vectors are extracted using the traditional UBM approach. The traditional UBM-based approach uses the same MFCC features to obtain the class alignment~(frame posteriors) and compute the sufficient statistics. These sufficient statistics are used to extract 600 dimension i-vectors using total variability matrix, $\textbf{T}$.

After the i-vector mean subtraction and length normalization, PLDA scoring calculated using batch likelihood ration which requires within-class~($\textbf{WC}$) and across class~($\textbf{AC}$) matrix. Within-class~($\textbf{WC}$) matrix characterizes how i-vectors vary from a single speaker and across class~($\textbf{AC}$) matrix characterizes how i-vectors vary between different speakers. The hypoparameters, $\textbf{WC}$ and $\textbf{AC}$, are estimated using in-domain NIST dataset.  
\section{DNN-senone-based GPLDA system}\label{sec:DNN senone posterior based GPLDA system}
In the previous section, a GMM is used to calculate the sufficient statistics for i-vector extraction. Though GMM is related to linguistic content, no knowledge of the linguistic content of the training speech is known. Recent success with DNN-senone for speaker recognition have shown that the sufficient statistics can be accumulated based upon supervised DNN senone posteriors~\cite{McLarenLeiFerrer,SnyderGarcia-RomeroPovey2015}.

Figure~\ref{fig:subfigureExample}~(b) illustrates how the i-vectors are extracted using a DNN-senone posterior approach. The DNN senone posterior approach uses automatic speech recognition~(ASR) features to compute the class alignments~(DNN posteriors) and speaker features are used for sufficient statistic estimation. The DNN parameters are trained using a transcribed training set. The supervised approach allows the DNN to provide phonetically-aware alignments. Once the sufficient statistics are accumulated, the i-vector extraction is performed in the same way as in Section~\ref{sec:GMM-UBM based GPLDA system}. After the development and evaluation data i-vector features are extracted, GPLDA modeling and scoring can be performed identically to the GMM-UBM based GPLDA approach.
\section{DNN senone training}\label{sec:DNN training}
The system is based on the multisplice time delay neural network~(TDNN) provided as a recommended recipe in the Kaldi toolkit for large-scale speech recognition~\cite{SnyderGarcia-RomeroPovey2015}. 40 MFCC features with a frame-length of 25ms was used for the ASR features. Cepstral mean subtraction~(CMS) was performed over a window of 6 seconds. The TDNN has six layers, and a splicing configuration was used as described in~\cite{PeddintiPoveyKhudanpur2015}. At layer 0, frames \{-2, -1, 0, 1, 2\} are spliced together. At layer 1, 3, and 4, frames \{-2, -1, 0, 1\}, \{-3, -2, -1, 0, 1, 2, 3\} and \{-7, -6, -5, -4, -3, -2, -1, 0, 1, 2\} are spliced together. The hidden layers use the p-norm (where p = 2) activation function. The hidden layers have input dimension of 300 and an output dimension 3500. The softmax output layer computes posteriors for 5346 triphone states.
\section{Speaker Recognition Methodology}\label{sec:Methodology}
The proposed methods were evaluated using the NIST 2010 SRE corpora coreext-coreext telephone-telephone condition~\cite{NIST2010}. The front-end consists of 20 MFCCs and delta and delta-delta were appended to create 60 dimensional frame-level feature vectors. The features are mean-normalized over a 3 second window. The energy-based voice activity detection~(VAD) was used to extract only voiced portion of speech utterances.

For the GMM-UBM PLDA approach, 2048 components UBM was trained using 32000 utterances which consists of SWB2 phase 2, SWB phase3, SWB cellular2, NIST 2004, 2005, 2006, 2008 data. A 600 dimensional total-variability space was also trained using all the utterances from SWB2 phase 2, SWB phase3, SWB cellular2, NIST 2004, 2005, 2006, 2008. Time delay neural network was trained using about 300 hours of English portion of Fisher data~\cite{cieri2004fisher}. All the NIST 2004, 2005, 2006 and 2008 data were used for GPLDA training. In this paper, all the experiments were trained and evaluated using Kaldi toolkit~\cite{PoveyGhoshalBoulianneEtAl2011}.

For short utterance speaker recognition experiments, truncation was done before the VAD and prior to the truncation, the first 10 seconds of active speech were removed from all utterances to avoid capturing similar introductory statements across multiple utterances. From the VAD outputs, it was found that around 50\% of frames are unvoiced, and the active speech part of truncated NIST 2010 coreext-coreext is around 15sec-15sec.

%
%
\begin{table}
\begin{center}		
\caption{\label{tab:standard nist results}\emph{Comparison of EER values of DNN-senone-based GPLDA and GMM-UBM GPLDA systems on the NIST 2010 SRE corext-coreext and 10sec-10sec conditions.}}
\scalebox{1.0}{
\begin{tabular}{l c c c} \hline
\textbf{System} & \textbf{Male} & \textbf{Female} & \textbf{Pooled} \\ \hline
\multicolumn{4}{l}{\textbf{NIST2010 coreext-coreext condition}} \\ \hline
\footnotesize GMM-UBM GPLDA & 1.53\% & 2.40\% & 2.16\% \\
\footnotesize DNN-senone GPLDA & \textbf{0.72\%} & \textbf{1.32\%} & \textbf{1.05\%} \\ \hline \hline
\multicolumn{4}{l}{\textbf{NIST2010 10sec-10sec condition}} \\ \hline
\footnotesize GMM-UBM GPLDA & 11.83\% & 13.73\% & 13.19\%  \\
\footnotesize DNN-senone GPLDA & \textbf{9.54\%} & \textbf{11.97\%} & \textbf{10.81\%} \\  \hline \hline
\end{tabular}}
\end{center}
\end{table}

\section{Results and Discussions}\label{sec:Results and Discussions}
%
Initial experiments were carried out to compare the performance of GMM-UBM GPLDA system and DNN-senone-based GPLDA system with standard NIST conditions. A second set of experiments were carried to compare the performance of the GMM-UBM GPLDA system and DNN-senone-based GPLDA system with truncated training and testing utterances, full-length training and truncated testing utterances. Both system's performance was also analyzed short utterances when GPLDA was trained using full-length~(2 min) and short-length utterances~(30 sec).
\begin{table}
\caption{\label{tab:15sec-15sec}\emph{Performance comparison of GMM-UBM GPLDA and DNN-senone GPLDA systems on NIST 2010 truncated 15sec-15sec evaluation condition when GPLDA is trained using full-length~(2 min) and short-length~(30 sec) GPLDA training data. The best performing systems by EER is highlighted across each row.}}
\begin{center}
\subfigure[\emph{GMM-UBM GPLDA}]{
\scalebox{1.0}{
\begin{tabular}{l c c c} \hline
\textbf{GPLDA training} & \textbf{Male} & \textbf{Female} & \textbf{Pooled} \\ \hline
\footnotesize Full-length & 15.36\% & 17.45\% & 16.87\% \\
\footnotesize 30 sec & \textbf{15.19\%} &\textbf{ 16.75\%} & \textbf{16.12\%} \\ \hline
\end{tabular} }}
\label{tbl:A}
\end{center}
				
\begin{center}
\subfigure[\emph{DNN-senone GPLDA}]{
\scalebox{1.0}{
\begin{tabular}{l c c c} \hline
\textbf{GPLDA training} & \textbf{Male} & \textbf{Female} & \textbf{Pooled} \\ \hline
\footnotesize Full-length & 12.73\% & 14.80\% & 14.64\% \\ 
\footnotesize 30 sec &  \textbf{12.21\%} & \textbf{13.91\%} & \textbf{13.47\%} \\ \hline
\end{tabular} }}
\label{tbl:B}
\end{center}
\end{table}

Table~\ref{tab:standard nist results} presents results comparing the EER values of DNN-senone-based GPLDA system against GMM-UBM GPLDA system on the standard NIST SRE 2010 coreext-coreext and NIST 2010 10sec-10sec conditions. It can be clearly observed from the Table~\ref{tab:standard nist results} that DNN-senone-based GPLDA system achieves above 50\% improvement in EER values over GMM-UBM based GPLDA system on NIST 2010 coreext-coreext male and pooled conditions. This is because DNN model provides phonetically-aware class alignments. However, when both GMM-UBM GPLDA and DNN-senone-based GPLDA systems were evaluated on NIST 2010 10sec-10sec evaluation condition, the performance improvement of DNN-senone GPLDA over GMM-UBM GPLDA system is reduced to 18\%. This is because the supervised DNN approach provides alignments based on phonetic content and when evaluation utterance length reduces, DNN-senone-based GPLDA fails to provide improvement as it has shown to long evaluation utterances.

\begin{table}[t]
\caption{\label{tab:full-15sec}\emph{Performance comparison of GMM-UBM GPLDA and DNN-senone GPLDA systems on NIST 2010 full-length~(2 min)-15sec evaluation condition when GPLDA is trained using full-length~(2 min) and short-length~(30 sec) GPLDA training data. The best performing systems by EER is highlighted across each row.}}
\begin{center}
\subfigure[\emph{GMM-UBM GPLDA}]{
\scalebox{1.0}{
\begin{tabular}{l c c c} \hline
\textbf{GPLDA training} & \textbf{Male} & \textbf{Female} & \textbf{Pooled} \\ \hline
\footnotesize Full-length & \textbf{6.75\%} & \textbf{9.21\%} & \textbf{8.61\%}  \\
\footnotesize 30 sec & 7.21\% & 9.69\% & 8.76\% \\ \hline
\end{tabular} }}
\label{tbl:A}
\end{center}
								
\begin{center}
\subfigure[\emph{DNN-senone GPLDA}]{
\scalebox{1.0}{
\begin{tabular}{l c c c} \hline
\textbf{GPLDA training} & \textbf{Male} & \textbf{Female} & \textbf{Pooled} \\ \hline
\footnotesize Full-length & \textbf{4.85\%} & 7.21\% & 6.32\% \\
\footnotesize 30 sec & 5.20\% & \textbf{6.96\%} & \textbf{6.14\%} \\ \hline
\end{tabular} }}
\label{tbl:B}
\end{center}
\end{table}

Tables~\ref{tab:15sec-15sec}~(a) and~(b) presents the results comparing the EER values of GMM-UBM GPLDA and DNN-senone-based GPLDA systems on NIST 2010 truncated 15sec-15sec evaluation condition when GPLDA is trained using full-length~(2 min) and short-length~(30 sec) GPLDA training data. It can be observed from Table~\ref{tab:15sec-15sec} that when the GPLDA model is trained on 30 sec utterances, DNN-senone-based GPLDA system achieves above 7\% improvements in EER values over full-length-based GPLDA training systems on pooled condition. This is because short length development i-vectors have speaker, session and phonetic variation and GPLDA is able to robustly model those variations. Further, the DNN-senone GPLDA achieves above 16\% improvement over GMM-UBM GPLDA systems when GPLDA model is respectively trained on short-length development data.

For several real world applications, longer utterances~(2min) can be used for enrollment and shorter utterances~(15sec) are required for verification. Tables~\ref{tab:full-15sec}~(a) and~(b) presents the results comparing the EER values of GMM-UBM GPLDA and DNN-senone GPLDA systems on NIST 2010 full-length-15sec evaluation condition. It can also be observed from Table~\ref{tab:15sec-15sec} and~\ref{tab:full-15sec} that DNN-senone GPLDA system achieves above 26\% improvement over GMM-UBM GPLDA system on full-length-15sec condition. These results suggest that though DNN senone systems failed show significant improvement on truncated enrolment and verification conditions, it shows significant improvement with full-length enrolment and truncated verification conditions. It was also observed from Tables~\ref{tab:full-15sec}~(a) and~(b) that when GPLDA model is trained on short-length development data instead of full-length development data, both systems failed to show improvement on full-length-15sec evaluation conditions. This because though GPLDA can robustly model speaker, and session and phonetic variability, full-length enrollment data only have speaker and session variations.
\section{Conclusion}\label{sec:Conclusion}
This paper investigated the effects of limited speech data in the context of speaker verification using DNN approach. The experimental studies have found that DNN-senone-based Gaussian probabilistic linear discriminant analysis~(GPLDA) system respectively achieved above 50\% and 18\% improvements in EER values over GMM-UBM GPLDA system on NIST 2010 coreext-coreext and truncated 15sec-15sec evaluation conditions. This is because the supervised DNN approach provides alignments based on phonetic content and it showed significant improvement with long utterance evaluation data. Further, when GPLDA model was trained on short-length utterances~(30sec) rather than full-length utterances~(2min), DNN-senone GPLDA system achieved above 7\% improvement in EER values on truncated 15sec-15sec condition. We believe that this is due to the fact that short length development i-vectors have speaker, session and phonetic variation and GPLDA is able to robustly model those variations. For several real world applications, longer utterances~(2min) can be used for enrollment and shorter utterances~(15sec) are required for verification, and in those conditions, DNN-senone GPLDA system achieved above 26\% improvement in EER values over GMM-UBM GPLDA systems. We believe that this is an important result for practical applications. In future work, BN features and DNN senone combined GPLDA systems will be investigated on short utterances.
\section{Acknowledgments}
This project was supported by an Australian Research Council (ARC) Linkage grant LP130100110. The authors would like to thank the Kaldi toolkit development team for their assistance.
									
\bibliographystyle{ieeetr}									
\bibliography{research}			
\end{document}